# Nonlinear Dynamic Behavior of an Oscillating Tip-Microlever System and Contrast at the Atomic Scale

J. P. Aimé,[1,*] R. Boisgard,[2] L. Nony,' and G. Couturier[1]

[1]*CPMOH Université Bordeaux 1, 351 Cours de la Libération, 33405 Talence, France*
[2]*IFE-SEFICESP, 40 rue Lamartine, F-33400 Talence, France*

In this paper the dynamic behavior of an oscillating tip-microlever system at the proximity of a surface is discussed. We show that the nonlinear behavior of the oscillator is able to explain the high sensitivity of the oscillating tip microlever and the observed shifts of the resonance frequency as a function of the tip surface distance without the need of introducing a particular short range force.
PUBLISHED IN Phys. Rev. Lett. 82(17), 3388-3391 (1999).

A few years ago, the resonant noncontact mode (hereafter noted NC-AFM) was developed to map tip surface interaction as shifts of the resonance frequency [1,2]. Since then, numerous experimental efforts have been performed showing that measurements of shifts were able to produce images at the atomic scale [3–7]. The "routine" achievement of the atomic resolution with the NC-AFM [7] is thought of as a real breakthrough in the field of the scanning probe microscopy. Moreover, the use of a high vibrating amplitude of the cantilever (CL) with the NC-AFM and the development of the intermittent contact (the so-called tapping mode) [8–10] have boosted theoretical works dedicated to a nonlinear analysis [11–14].

Before going a step further, it is worth discussing general ideas at a qualitative level. Leaving aside the technical point that the use of a high amplitude reduces the influence of the fluctuations in frequency [2], a first, counterintuitive, result is the use of a large vibrating amplitude allowing the atomic resolution to be achieved. Typically, the amplitudes used are 100 times greater than the vertical motion of the surface required, maintaining a given value of the frequency shift. The most common and shared idea is that to probe an attractive field with an oscillator requires the use of a small amplitude in order to keep the force field nearly constant throughout the vibrating amplitude. The idea is that shifts in resonance frequency are uniquely due to the gradient force variations.

An image with the atomic resolution based on a repulsive interaction appears quite easily understandable, but getting the same resolution when the attractive van der Waals forces are involved is very puzzling. The van der Waals atom-atom interaction and the finite size of the interacting objects lead to a tip-sample interaction smoothly varying as a function of the CL-surface distance. For instance, the sphere-plan interaction leads to an attractive force with a $d^{-2}$ power law. The force gradient dependence, with a power law $d^{-3}$, gives a too smooth variation of the resonance frequency and is unable to predict the observed shape of the frequency changes as a function of the tip-sample distance.

The key point with an AFM is that the recorded information comes from the tip-sample interaction, but when the CL is externally periodically excited, the recorded image is a measure of the oscillating behavior of the CL. When the tip-microlever system experiences an attractive field, the oscillator can exhibit a nonlinear behavior [14–16]. For an oscillator set at a drive frequency slightly below the resonance one, a bifurcation from a monostable to a bistable state can occur as soon as the vibrating amplitude is large enough [14]. Keeping the amplitude constant, the shifts in resonance frequency show a continuous, monotonous, variation as a function of the tip surface distance that can be very large when nonlinear effects take place, and in turn amplify small changes of the strength of the interaction.

*Variational method, principle of least action.*—A variational method can be developed based on the principle of least action [14]. The action $S[(x)]$ is a functional of the path $x(t)$ and is extremal between two fixed instants.

$$S[x(t)] = \int_{t_a}^{t_b} L(x, \dot{x}, t)\,dt, \quad (1)$$

where $L$ is the Lagrangian of the system. The main aim of the use of the variational principle is to employ a trial function that allows us to perform a nonperturbative analytical treatment in which the dissipation is included. As for the treatment of the Duffing oscillator [15,16], we focus on the behavior of the harmonic solution of the form $x(t) = A(D)\cos[\omega t - \phi(D)]$. Using a sphere-plan interaction [17], a Lagrangian is obtained,

$$\begin{aligned}L &= T - U + W \\ &= \frac{1}{2}m\dot{x}^2 - \left(\frac{1}{2}kx^2 - xf\cos(\omega t) - \frac{HR}{6(D-x)}\right) \\ &\quad - \gamma x\dot{x},\end{aligned} \quad (2)$$

where $H$ is the Hamaker constant, $R$ the radius of the tip, $D$ the distance between the surface and the CL at the equilibrium position at rest, and $\gamma$ the damping coefficient. The parameters of the path are $A$ and $\phi$, and the variational principle $\delta S = 0$ becomes a set of two partial differential equations,

$$\frac{\partial S}{\partial A} = 0,$$

$$\frac{\partial S}{\partial \phi} = 0.$$

After some tedious calculations, two coupled nonlinear equations are obtained,

$$\cos\phi = Qa(1 - u^2) - \frac{\alpha}{3}\frac{a}{(d^2 - a^2)^{3/2}}, \quad (3a)$$

$$\sin\phi = au, \quad (3b)$$

where the amplitude, the distance, and the frequency are expressed in reduced coordinates $a = A/A_0$ and $d = D/A_0$, $u = \omega/\omega_0$, where $\omega_0$ and $A_0$ are the frequency and amplitude at the resonance far from the surface. Finally, the dimensionless parameter $\alpha$ is used,

$$\alpha = \frac{HRQ}{k_c A_0^3}, \quad (4)$$

where $k_c = m\omega_0^2$ is the CL stiffness. At a fixed drive frequency, with $\alpha < 1$ a bifurcation from a monostable to a bistable state can be observed, while for $\alpha > 1$, the instabilities disappear [14]. When a molecule-plan interaction is used, the dimensionless parameter becomes $\alpha = \frac{HR^3Q}{k_c A_0^5}$ and the power law is $(d^2 - a^2)^{-7/2}$ [18].

Solving Eqs. (3a) and (3b) gives the relationships between the amplitude, the distance, and the frequency,

$$d_{\pm} = \sqrt{a^2 + \left(\frac{\alpha}{3[Q(1-u^2) \mp \sqrt{1/a^2 - u^2}]}\right)^{2/3}} \quad (5)$$

from which is derived the resonance curve at a given distance $d$ (Fig. 1).

$$u = \sqrt{\frac{1}{a^2} - \left[\frac{1}{2Q}\left(1 \pm \sqrt{1 - 4Q^2\left(1 - \frac{1}{a^2} - \frac{\alpha}{3Q(d^2 - a^2)^{3/2}}\right)}\right)\right]^2}. \quad (6)$$

In a NC-AFM experiment the amplitude is kept constant at the resonance value $A_0$ at any distance from the surface, and this experimental condition is obtained by setting $a = 1$ in Eq. (6). The shift of the resonance frequency as a function of the distance $d$ is given by the branch $d_+$,

$$1 - u = \frac{\Delta\nu}{\nu_0} = 1 - \sqrt{1 - \left(\frac{1 + \sqrt{1 + 4Q\alpha/3(d^2-1)^{3/2}}}{2Q}\right)^2}. \quad (7)$$

Here it is useful to use an approximation allowing Eq. (7) to be more tractable. Let us consider the case corresponding to

$$Q \gg \frac{1}{\sqrt{1 - u^2}}. \quad (8)$$

The inequality (8) means that the relative width of the resonance curve is smaller than that of the relative shift of the resonance frequency. For the large values of the amplitude $A_0$, a further simplification can be done and Eq. (7) gives the frequency shift

$$\frac{\Delta\nu}{\nu_0} \sim \frac{HR}{6k_c A_0^3 (d^2 - 1)^{3/2}}. \quad (9)$$

Equation (9) indicates that a high sensitivity can be obtained for $d$ values close to 1. Practically, the use of Eq. (9) [or Eq. (7)] to fit the experimental results might help to decide whether or not the tip touches the surface.

Equation (9) gives a result similar to the one obtained with a perturbative approach, which does not consider the dissipation [19]. Equation (13) of Ref. [19] predicts that $\Delta\nu$ scales as $A_0^{-3/2}$, a power-law dependence which can be derived from Eq. (9) with the assumption that $A_0$ is large and $D$ is nearly equal to $A_0$. Writing $(d^2 - 1)^{3/2} = A_0^{-3}(D^2 - A_0^2)^{3/2}$ for $D = A_0 + \varepsilon$ with $\varepsilon \ll A_0$ one gets $\Delta\nu \sim A_0^{-3/2}\varepsilon^{-3/2}$, the validity of this expression being a function of the ratio $\varepsilon/A_0$. To measure $\Delta\nu(A_0)$, the tip-sample distance $\varepsilon$ has to be kept constant while varying the amplitude $A_0$. As discussed in Ref. [19], the work performed by Kitamura and Iwatsuki [5] gives a $\Delta\nu$ of 10 Hz at $A_0 = 10$ nm and 150 Hz at $A_0 = 1.5$ nm. Thus a ratio $\ln[\nu(A_0 = 1.5)/\Delta\nu(A_0 = 10)]/\ln[(A_0 = 1.5)/(A_0 = 10)] = -1.43$; this result shows the influence of the approximation as $A_0 = 1.5$ nm is a rather small value of the amplitude. To evaluate the distance $\varepsilon$, we use the value calculated in Ref. [20] that gives the tip surface distance at which a force image can be recorded in a pure attractive regime between a diamond tip and a NaCl surface. Tang *et al.* found a tip-sample distance $\varepsilon$ of about 0.35 nm. Inserting this value in Eq. (9), a linear fit of the log log plot of Log($\Delta\nu$) versus log($A_0$) gives a slope of $-1.438$ between $A_0 = 1$ nm and $A_0 = 15$ nm and $-1.464$ for resonance amplitudes ranging between 1 and 40 nm, thus a remarkable agreement with the experimental data.

Because of the use of reduced coordinates, the power-law dependence does not appear clearly from Eq. (9). The use of reduced coordinates is an usual way to describe the stationary state of an oscillator. But while with a linear behavior the amplitude of the forcing can be scaled out, when a nonlinear behavior occurs, the amplitude of the forcing becomes a new operative parameter that can significantly alter the dynamical phenomena exhibited by the oscillator. Nevertheless, it has been demonstrated that the use of reduced coordinates is particularly fruitful to describe experiments in which a fixed driven frequency is chosen [14,21]. The main interest is to exhibit a dimensionless parameter $\alpha$ [Eq. (4)] being able to give the conditions for which the bifurcation of the amplitude and the phase are observed. The predictions given by the model were experimentally verified [14,21].

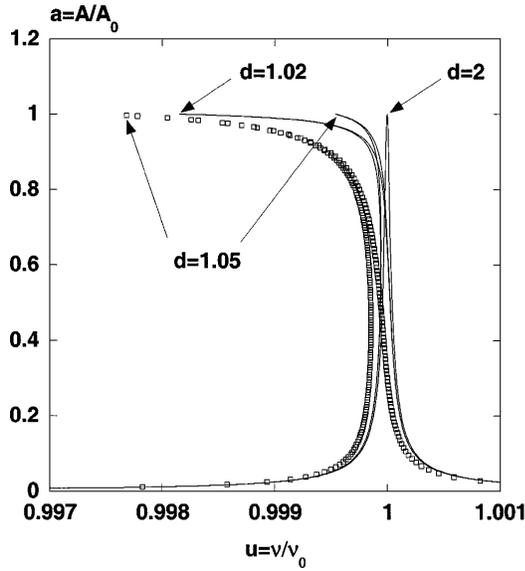

FIG. 1. Resonance curves calculated with Eq. (6) for different CL-surface distances $d$, with $Q = 20\,000$, $HR = 10^{-27}$ J m, and $A_0 = 10$ nm (continuous line $k_c = 11$ N m$^{-1}$). The flatness of the distorted curve gives a slope $da/d\nu \sim 0$, which means that any small fluctuations of the amplitude can lead to a large variation of the measured frequency (see text and Fig. 2). Also shown is the influence of the cantilever stiffness on the shape of the resonance curve (empty squares $k_c = 2$ N m$^{-1}$).

For the present purpose, the structure of Eq. (6) suggests that rather than $\alpha$ it is more suitable to seek for critical values of the ratio $\alpha/Q$. The location of the bifurcation is given by the condition $du/da = 0$ (point $C$ in Fig. 2). The resonance peak starts to distort when all the points $A$, $B$, $C$, and $D$ collapse in two points. For a given $u$ value two amplitudes are reached, $a(u)$ and $a = 1$. Using this property we derived an analytical expression from Eq. (7) [22] giving the domain of bifurcation as a function of $\alpha/Q$ and $d$ (Fig. 3). The curve determines the frontier between two domains, the domain in which the resonance peak is slightly distorted and keeps approximately a Lorentzian shape and the domain of bifurcation given by the upper part of the curve. Note that the domain of bifurcation is significantly reduced for low values of $Q$. This result illustrates a more general principle that the domain of bifurcation and, if it occurs, a chaotic behavior are less important with maximum dissipation [23].

As soon as the quality factor is large enough, the sensitivity of the oscillating CL is given by the ratio $HR/k_c A_0^3$. For a given value of the product $HR$, the same variation of the frequency as a function of the reduced distance is expected when the product $k_c A_0^3$ is kept constant. For example, for a $k_c$ value 8 times smaller, to get the same frequency shift at the same reduced distance requires the use of an amplitude $A_0$ twice larger. In other words, one gets the same sensitivity with a distance $D$ and a closest distance $\varepsilon$ twice larger. This result can be directly derived from Eqs. (7) or (9).

The description of the instabilities experienced by the CL dynamical system is not straightforward. When the CL oscillates, the oscillating behavior is described with two coordinates in the phase space. A consequence is that the phase relationship between the oscillator and the attractive force varies as a function of the CL surface distance, and experimental situations were found where the attractive interaction acts as a repulsive one [14]. In NC-AFM the amplitude is kept constant and equal to the resonance one; thus the phase should normally be a constant. But as shown in Figs. 1 and 2, the ability to follow the resonance frequency becomes more and more difficult as the distortion of the resonance curve increases. The fluctuations of the amplitude, even of the order of $\delta A_0/A_0 = 10^{-4}$, can induce a measured frequency different than that of the resonance one, thus giving a value of the phase different from $-\pi/2$ (Fig. 2). Even worse is the fact that the oscillator has the possibility of being in a state belonging to the lower instable curve producing a phase jump.

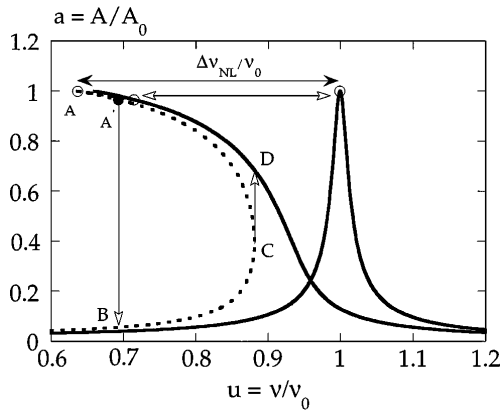

FIG. 2. Shifts in resonance frequency $\Delta\nu_{NL}$ correspond to shifts given by the black arrows. Because of the experimental uncertainties and the flatness of the distorted curve (Fig. 1), even with a constant amplitude mode various situations can occur. If the oscillator is trapped on the lower instable curve (black circle on the dotted line), then a cycle of hysteresis following the path $A'BCD$ follows.

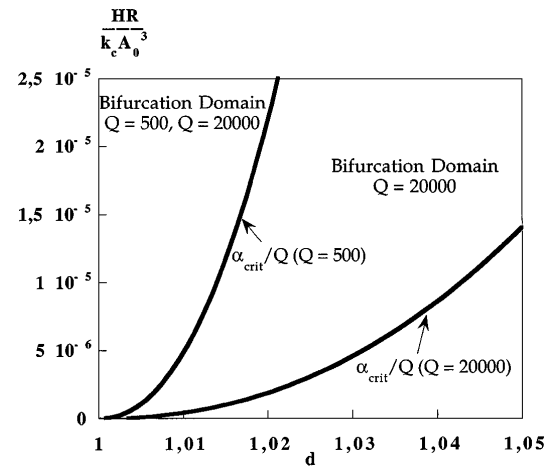

FIG. 3. Determination of the domain of bifurcation calculated for two values of the quality factor $Q$. The curves are calculated with the equation given in Ref. [22]. The curves $\alpha_{\rm crit}/Q$ give the frontier above which the resonance peak starts to distort.

While Eqs. (7) and (9) give qualitatively a similar behavior to the ones observed experimentally [3,5–7], quantitatively the use of such analytical expressions remains questionable. As shown by various groups, spectacular variations on the image were observed as a function of the chosen frequency shift $\Delta \nu$ [7]. The larger the values of $\Delta \nu$, the smaller is the tip-sample distance at which the images are recorded. Therefore as a function of the tip-sample distance, the surface can interact in a different manner with the tip. Not only the geometrical factors may vary for different values of the frequency shift, making the whole dependence of $\Delta \nu$ a mixing of different power laws as a function of the tip-sample distance [18], but also, particularly when the resonance curve are strongly distorted (Figs. 1 and 2), one cannot be completely sure that the resonance frequency is measured.

In conclusion, the present Letter aims to describe some of the evolution of the vibrating tip-microlever system. Here we focus on shifts of the resonance frequency that have been shown to be able to record images with the atomic resolution. The theoretical description is able to predict the variation of the resonance frequency as a function of the tip-surface distance. It also shows that a linear analysis cannot account for the frequency shift variation and thus will be unable to explain the large change of the resonance frequency when the tip-sample distance is slightly varied.

This paper also shows that there is no need to introduce particular short range forces to explain the way the shifts of the resonance frequency is varying as a function of the tip-sample distance. In other words, depending on the surface investigated, the observed contrast at the atomic scale could be an image of tiny variations of the fluctuating forces. Also we show that because of the high-$Q$ values used in the NC-AFM experiments distortion of the resonant peak always occurs.

We have benefited from many discussions with the participants of the Workshop NC-AFM '98. Particularly, it is a pleasure for one of us to thank Dr. A. Schwarz for many illuminating discussions and Dr. F. Giessibl for giving us a preprint prior to publication.